\documentstyle[aps,preprint]{revtex}

\begin{document}

\author{
Daniele Passerone$^{a,b,c}$\cite{corrauth}, Ugo Tartaglino$^{b,c}$\cite{aff0},
 Furio Ercolessi$^{b,c}$\cite{aff}, and Erio Tosatti$^{b,c,d}$\cite{aff2} \\
{\it a) Max-Planck-Institut f\"ur Festk\"orperforschung, Stuttgart, Deutschland}\\
{\it b) Istituto Nazionale per la Fisica della Materia (INFM)}\\
{\it c) International School for Advanced Studies, Trieste, Italy}\\
{\it d) International Center for Theoretical Physics,}\\
{\it Trieste, Italy}
}
\title
{Surface molecular dynamics simulation with two orthogonal surface steps:
how to beat the particle conservation problem}
\maketitle
\begin{abstract}
Due to particle conservation, Canonical Molecular Dynamics (MD) 
simulations fail in the description of surface phase 
transitions involving coverage or lateral density changes. However, 
a step on the surface can act effectively as a source or a sink 
of atoms, in the simulation as well as in real life. A single 
surface step can be introduced by suitably modifying planar 
Periodic Boundary Conditions (PBC), to accommodate the generally 
inequivalent stacking of two adjacent layers. We discuss 
here how, through the introduction of {\em two orthogonal} surface 
steps, particle number conservation may no longer represent 
a fatal constraint for the study of these surface transitions. 
As an example, we apply the method for estimating 
temperature-induced lateral density increase of the reconstructed
 Au (001) surface; the resulting anisotropic cell change is 
consistent with experimental observations.
Moreover, we implement this kind of scheme in conjunction with the 
variable curvature MD method, recently introduced by our group. 

\end{abstract}

\section{introduction}

Classical Molecular Dynamics (MD) simulation of crystal surfaces is  
generally inadequate for the description of physical situations 
involving large density and structure changes including some surface 
reconstructions, incomplete melting, thermal preroughening and 
roughening, etc. due to the constraint of particle conservation. 
A single monatomic step can act as a source or sink of 
atoms: given a request for lateral density variation the step
can move normally to its direction, to accommodate to it.
%OLD:
%Pairs of steps of opposite signs are generally easy to implement 
%NEW:
Pairs of steps of opposite sign are generally easy to implement 
%--
in a slab simulation: but this is not optimal, for it leads
to a very limited size terrace, with non-negligible step-step 
interactions ($\simeq 1/L^2$, if $L$ is the terrace size), and 
general hindering of the step movement. 
To minimize this problem, one can introduce a {\em single} step 
per unit cell, which will weakly interact only with its own replicas 
($L$ being in this case the entire lateral size 
of the simulation cell) through periodic boundary conditions (PBC). To 
this end, the PBC must be modified so that a layer $n$ on the cell 
right hand side is joined at the left hand side not with itself, but 
with layer $n+1$. In this manner the single step is introduced
topologically. That scheme was applied with success
to the simulation of the lateral density increase of the 
reconstructed  hexagonal overlayer of Au$(001)$ \cite{001}.
The movement of the step was indeed observed, with an enhanced
mobility connected in that case with the incommensurability between
first and second layer \cite{single}.

However, a lateral density adjustment will in general take place
in both directions on the surface and not just along one. 
The impossibility for the 
density to re-adjust in both directions (as implied by the unidirectional 
single step construction, and by rigid overall PBC parallel to the step) 
can also hinder interplanar sliding which might in real life require
simultaneous sliding in both planar directions.
The natural solution to this problem is to allow the same freedom 
in both orthogonal surface directions. A practical realization of this 
task is presented here through the implementation of two orthogonal steps.
%OLD:
%We will first of all examine the technical difficulties of  
%NEW:
We will first of all examine the practical
%--
implementation and the applicability to surfaces of unbent as well
as of bent crystals, in the light of the recently introduced 
variable curvature molecular dynamics method \cite{PRBCURV}.
We will then present direct examples of application, exploiting
%OLD:
%as usual gold surfaces as a test case.
%NEW:
as done previously \cite{single,PRBCURV} gold surfaces as a test case.
%--

\section{The double step}

A surface with regularly repeated steps separated by large terraces
%OLD:
%is a vicinal surface. A method for obtaining a vicinal surface
%in simulation is to rotate a slab by an angle $\theta = \arctan (L/h)$
%NEW:
is a vicinal surface. A method for generating a vicinal surface
for simulation is to rotate a slab by an angle $\theta = \arctan (L/h)$
%--
with a flat surface and dimensions $L_x \times L_y \times L_z$
around the step axis (called $x$) (with $L$ the
distance between the steps and $h$ the 
interplanar separation) and to ``cleave'' the crystal surface 
accordingly, generating a sequence of monatomic steps.
If $L$ is chosen equal to $L_y$, a single step is produced. In a typical
high symmetry direction of an fcc crystal, such as $(111)$ or $(001)$,
the stacking of the layers is $ABCABC...$ or $ABABAB...$. If a single
step is produced, the portion of terrace above the step and the one below
will belong to layers of different sublattices, thus causing a 
mismatch in the periodic boundary conditions (PBC).
This problem is solved through the introduction of an
additional translation vector in the PBC, absorbing 
the mismatch between the two different layers \cite{single}. An 
equivalent solution is to modify the dimensions and orientation 
of the simulation cell, which will no longer be built with an integer
number of unit cells nor with walls parallel to high-symmetry planes
of the crystal \cite{Passerone}. We prefer here the former approach.

If we now wish to add a single step along the orthogonal $y$ 
direction as well, a problem in the PBC will obviously appear 
also in the $x$ direction, and we will need {\em two}
rotations of the slab. The first, through an angle 
$\theta\simeq L_y/h$ about the $x$ axis brings the $y$ axis into 
$y'$. The second, through an angle $\phi \simeq L_x/h$, about $y'$.

The dimension of the cell is chosen so that a particle and its 
image lie at the same height $z$. 

\section{Application to a flat surface}

As an example, we consider a Au$(001)$ surface, whose hexagonal
reconstructed layer is incommensurate, with a cell approximately 
$(n \times 5)$  with $n\simeq 30$.
In the $[110]$ (henceforth $y$) direction, 6 surface rows, shifted alternatingly so as
to form a quasi triangular layer, are stacked on 5 rows of the 
%OLD:
%underlying square second layer. In the $[1-10]$ (henceforth $x$) direction, one soliton-like 
%NEW:
underlying square second layer. In the $[1\overline{1}0]$
(henceforth $x$) direction, one soliton-like 
%--
extra atomic row is present every $n$ second layer rows.
The surface finally appears as 
in Fig. 1 , where a reconstructed Au$(001)$ surface, 20 layer 
thick, is shown. Although there is in reality a single large
terrace, that appears in the form of four sub-terraces of
height 1,0,0, -1. The corrugation given by the $(5 \times 1)$ 
%OLD:
%periodicity is clearly visible. 
%NEW:
reconstruction periodicity is clearly visible. 
%--

To demonstrate how the two orthogonal steps are going to act
as grand-canonical sources of surface atoms, we will exploit
one well established but rather peculiar property of this surface:
Au$(001)$  {\it increases} its lateral density upon heating 
%OLD:
%\cite{001,noi}. Physically this is due to the competing role of the 
%NEW:
\cite{addensamento,PasseroPhD,single}. Physically this is due to the competing role of the 
%--
$d$- and $s$-orbitals of gold: the consequence of this competition 
is a giant outwards relaxation, resulting in a consequent planar 
contraction of the first layer.

%OLD:
%Since the 6 over 5 commensurability pinning in the $y$ direction 
%is much stronger than the $n+1$ over $n$ pinning in
%NEW:
Since the (6 over 5) commensurability pinning in the $y$ direction 
is much stronger than the ($n+1$ over $n$) pinning in
%--
the $x$ direction, the lateral contractions upon heating will generally be
%OLD:
different between x and y. Our goal here will be to discover what
our approach predicts for them using the two-steps geometry. 
%NEW:
different between x and y. Our goal here will be to discover by simulation
what should happen in reality by using the two-step geometry. 
%--
We modeled the surface with a slab with 20 layers of 1800 
atoms each. The two surfaces were reconstructed with 31 rows 
over 30 in the $x$ direction and 6 rows over 5 in the $y$ direction.
Our interatomic potential was the glue model of Ercolessi {\it et al.}
\cite{solitacolla} with a many body term mimicking the metallic cohesion
of valence s and d electrons. Newton's equations were numerically 
integrated with standard algorithms, and velocity rescaling is used 
to control temperature. After an initial quench to relax the slab
%OLD:
%atomic positions, we heated slowly up to  $1000 K$. 
%NEW:
atomic positions, we heated slowly (rate 100 Kelvin/nsec) up to  $1000 K$. 
%--
Fig. 2 shows a top view of the surface of Fig. 1 before and after it was
heated in this way. 

%OLD:
%At the beginning, the four sub-terraces had equal dimensions (left side of the 
%figure).
%At $T=900 K$ (right side) the situation has now
%NEW:
At the beginning ($T=0$) the four sub-terraces had equal dimensions
(left panel).
At $T=900 K$ (right panel) the situation has now
%--
changed, with a visible withdrawal (even if superposed with 
meandering) of the x-step along the positive y direction, but no
visible shift of the y-step along the x direction. As a result,
the two lower subterraces 0, -1 have expanded at the expense of
the two upper ones 1, 0. The relative x-step withdrawal is about 4\%.
Since that happens in one direction only, this measures exactly
the relative lateral density increase in the top layer of our Au(001)
on heating model between zero and 900 K. This value equals very closely
the magnitude of the known surface lateral density increase of 
%OLD:
%real Au(001) from room temperature to 900 K \cite{addensamento}.
%NEW:
real Au(001) from room temperature to 900 K \cite{addensamento,PasseroPhD}.
%--
%OLD:
%Interestingly it is mostly the weakly pinned x-step that acted 
%NEW:
Not surprisingly it is mostly the weakly pinned x-step that acted 
%--
as a source of atoms for the terrace density increase, while hardly
%OLD:
%anything happened to the strongly pinned y-step. Thus the reconstruction
%went with temperature from $(5 \times 30)$ to roughly $(5 \times 28.5)$,
%NEW:
anything happened to the strongly pinned y-step. The reconstruction
periodicity drift with temperature from $(5 \times 30)$ to roughly $(5 \times 28.5)$,
%--
which is precisely the trend seen experimentally \cite{addensamento}.
 
\section{Double step and variable curvature}

%OLD:
%Another interesting application of the double step geometry 
%NEW:
A second interesting application of the double step geometry 
%--
is in conjunction with the Variable Curvature Molecular Dynamics 
(VCMD) method \cite{PRBCURV}. That method allows the simulation 
%OLD:
%of a curved slab, the radius of curvature treated as an additional 
%NEW:
of a bent slab, the radius of bending curvature treated as an additional 
%--
degree of freedom. In the force-free case, the slab will adopt
a spontaneous curvature in response to the surface stress difference 
of its two faces \cite{ibach}. Alternatively, a bending force or constraint
can be applied to the slab, to force a given curvature. The
%OLD:
%associated fields of strain and stress in the slab will in turn change
%NEW:
associated fields of strain and stress in the slab will change
%--
its overall free energy and in particular that of its
%OLD:
%surfaces \cite{ugo}. In response, the surface lateral density
%NEW:
surfaces \cite{ugo}. In response, the equilibrium surface lateral density
%--
may also generally change, and we can gauge that change with our double
step method. Special care must be taken in this case concerning 
PBC: if the slab is bent, a particle with velocity parallel to the 
bending direction which crosses
the boundary of the simulation cell, must remain on the same `orbit', thus the 
translation vector added to the usual PBC refolding vectors should be
also parallel to the bending direction.
With this in mind, we have considered two important cases under bending: 
%OLD:
%Au(001), and the Au(111) with its $(\sqrt 3 \times 23)$ reconstruction.
%NEW:
Au(001) $(1\times 5)$ and
the Au(111) with its $(\sqrt 3 \times n)$ reconstruction.
%--
Details of both calculations will be reported in a forthcoming paper 
%OLD:
%\cite{forseprl}. Here we will give some preliminary results concerning 
%Au$(111)$. This surface possesses the well known herringbone 
%reconstruction consisting in alternating $(\sqrt 3 \times 23)$ 
%domains with a mutual orientation of $120$ degrees, separated
%by a sort of soliton twin boundary.
%NEW:
\cite{forseprl}. Here we just give some preliminary results concerning 
Au$(111)$. 
%--
Under compressive strain, the density mismatch 
between the substrate and the overlayer decreases, and the $\sqrt 3 \times n$ periodicity
should change, in particular n should increase.

%OLD:
%We simulated a single reconstructed domain, with 6 parallel 
%solitons which are parallel to the $[11-2]$ (henceforth $x$) direction and two orthogonal steps, a total of about 1200 surface 
%atoms, for a slab thickness of 40 layers. 
%We applied a finite curvature in the $[1-10]$ (henceforth $y$) direction, and 
%NEW:
We started with a single $(\sqrt{3}\times 11)$ reconstructed domain
($n=11$ is the equilibrium value in the glue model \cite{bartol}),
with 6 solitons which are parallel to the $[11\overline{2}]$ (henceforth $x$)
direction and two orthogonal steps, a total of about 1200 surface 
atoms, for a slab thickness of 40 layers. 
We applied a finite curvature in the $[1\overline{1}0]$ (henceforth $y$)
direction, and 
%--
focused on the face under compression (see Fig. 3).

Fig. 4 shows the phase difference of the overlayer over the second layer
with and without bending strain.
The flat surface shows 6 solitons along the $x$ direction 
(the peak of one of the steps is visible), whereas under a 
compressive strain of about 4\% the 
%OLD:
%period of reconstruction has doubled, and there are only 3 solitons.
%NEW:
period of reconstruction has doubled to $(\sqrt{3}\times 22)$ and
there are only 3 solitons.
%--
The step parallel to the soliton advances in 
this case, accepting atoms from the terrace.
On the other hand, the other step (orthogonal to the solitons) 
%OLD:
%remains pinned due to strong commensurability.
%NEW:
remains pinned, most likely due to strong commensurability.
%--
Hence in comparison with the Au(001) studied in the previous
chapter, we have step motion parallel rather than orthogonal
to the soliton, and we also demonstrate step advancing rather than
withdrawing.
 
\section{Conclusions and perspectives}

Two orthogonal single steps can be built topologically
on a simulated crystal surface through periodic boundary conditions.
This device is shown to permit, under favorable conditions, 
an effectively grand canonical study of the surface with
molecular dynamics (MD) simulation, otherwise strictly canonical.
Any change of surface coverage will reflect in an advancement
or retraction of the two steps, that can be naturally quite
anisotropic. The method is shown to work by means of two case studies.
the thermal increase of lateral surface density of Au(001), and 
the strain-induced decrease of that of Au(111).

%OLD:
%In the first case only one step moved, but the presence of two orthogonal steps
%allows to separate the lateral expansion in two directions. The conclusion that 
%NEW:
In both cases only one step eventually moved, but the presence of two
orthogonal steps allowed to separate the lateral expansion in two directions.
The conclusion that 
%--
%OLD:
%allows to separate the lateral expansion in two directions. The conclusion that 
%the contraction is dominant in one direction with respect to the other could not have
%been drawn with a single step geometry.
%NEW:
the contraction is dominant in one direction with respect to the other,
althoug physically plausible, could not have
been drawn with a single step geometry.
%--

In the case of surfaces under strain, we are 
presently applying the same method to the curved Au $(100)$, and the results will appear in a future 
%OLD:
%paper \cite{forseprl}; we expect to observe (as suggested by preliminary 
%simulations) a density change in the
%direction of `long' periodicity in response
% to an external strain in the `$(6 \times 5)$' direction, and the 
%double step geometry will be crucial to signal and quantify this effect.
%NEW:
paper \cite{forseprl}.
%--

\begin{center}
 \textbf{ACKNOWLEDGMENTS}
\end{center}

We acknowledge support from MURST COFIN99, and from INFM/F.
Research of D. P. in Stuttgart is supported by the Alexander Von Humboldt 
foundation.

\section*{Figure captions}
\begin{itemize}
\item Figure 1: The 20-layers slab of the reconstructed Au (100) surface with two 
orthogonal steps; the corrugations given by the $5 \times 1$ periodicity is clearly visible.
\item Figure 2: Top view of the Au (100) surface before and after heating at $900 K$ (brighter atoms are higher, thus the top left 
terrace is the highest one). Before heating, the four subterraces have the same size (left);
after heating, the horizontal step has retracted, signalling a lateral density increase (right). The numbers (0,-1,1) indicate the relative height of the four subterraces.
\item Figure 3: First few layers of the simulated Au (111) slab under compressive strain along
the $1\overline{1}0$ ($y$) direction.
Both the two orthogonal steps and the externally imposed curvature can be observed($T=800 K$).
\item Figure 4: Change in periodicity (phase difference of the overlayer over the
second layer) of the Au (111) surface reconstruction under compressive 
strain: the flat surface shows 6 solitons along the $y$ direction (lower curve), whereas
under a compressive strain of about $4 \%$ the period of reconstruction has doubled (upper
curve).
\end{itemize}


\begin{thebibliography}{99}
\bibitem{Passerone}We found easier to produce new slabs in this second way; moreover, if bending slabs are simulated,
 with this choice the high symmetry directions of the crystal are parallel 
to the bending direction of the cell.
\end{thebibliography}
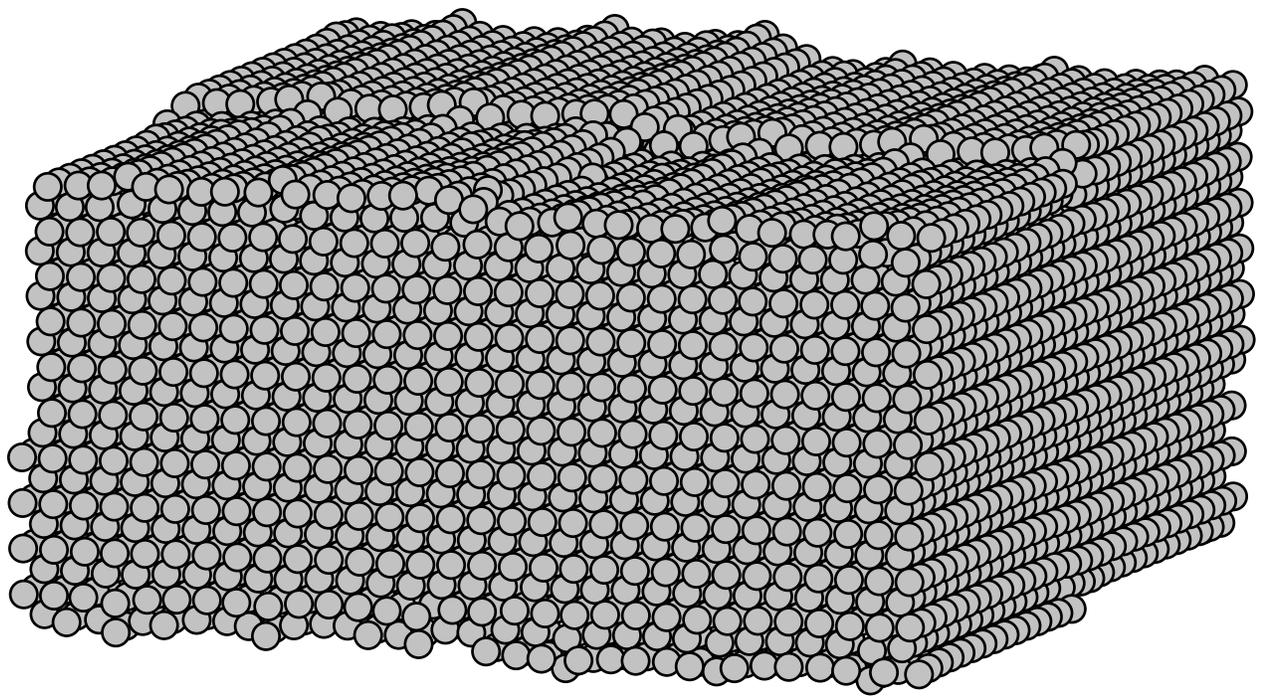
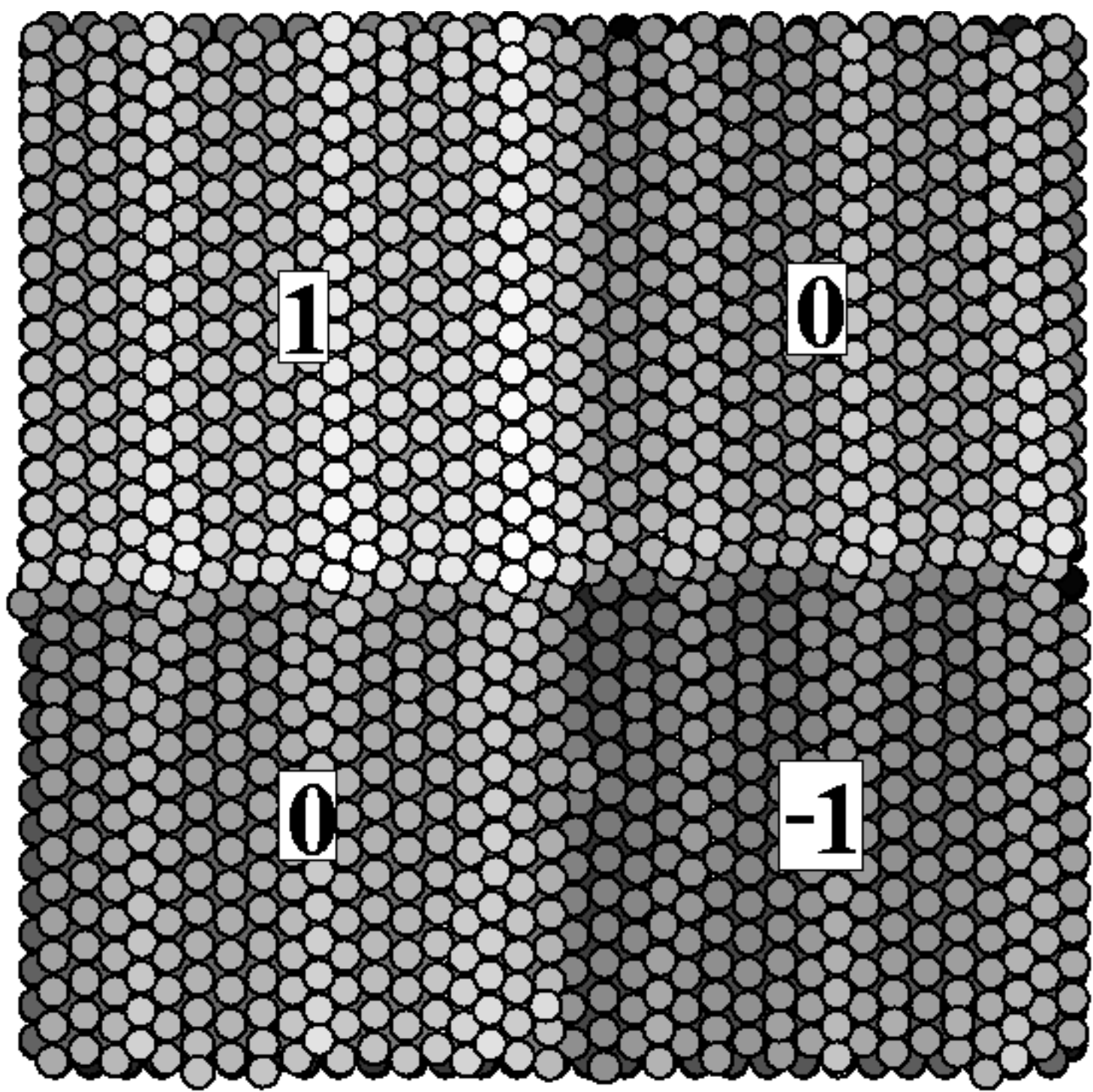
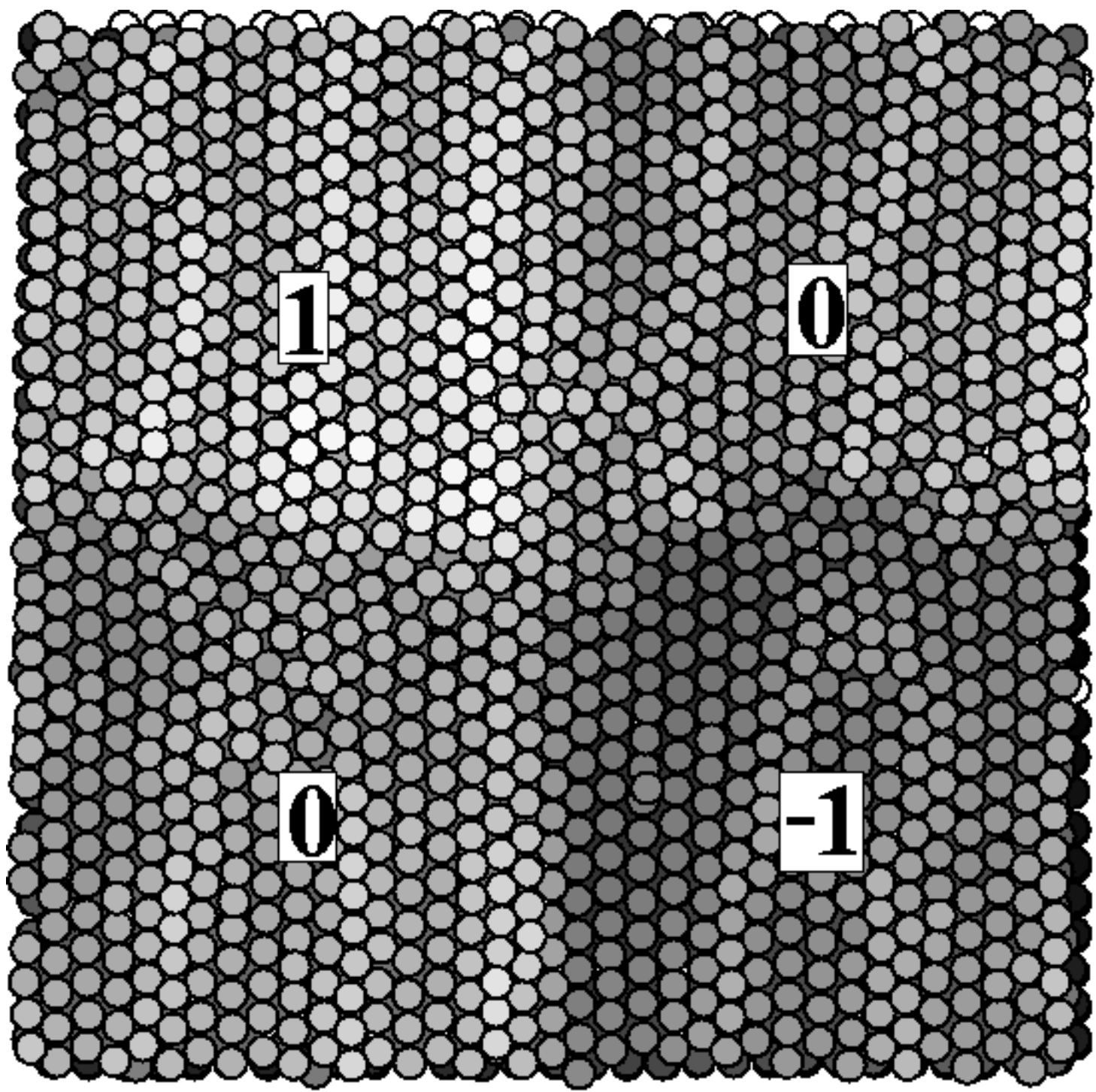
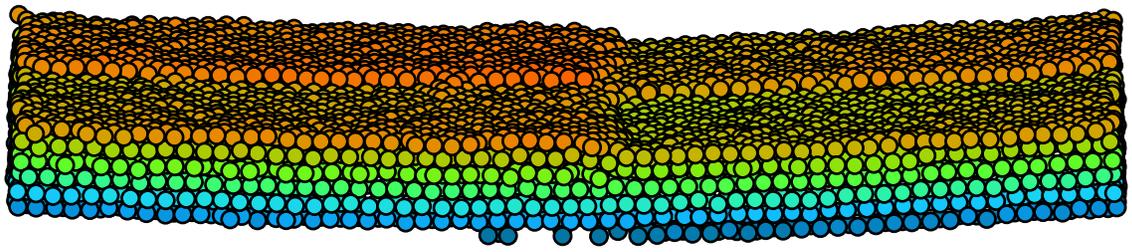
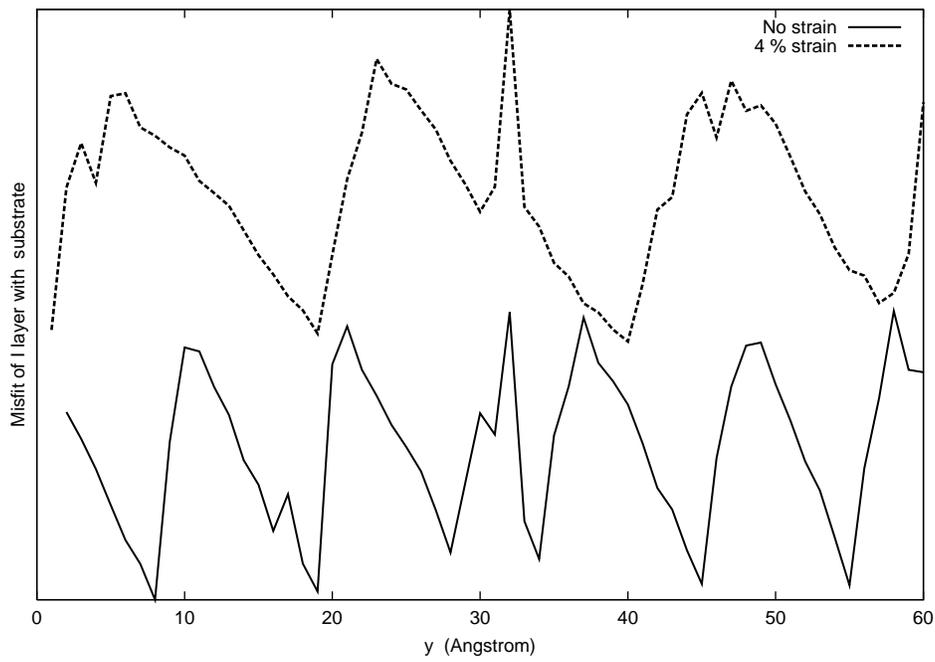
\end{document}